\documentclass[a4paper,twocolumn]{revtex4}

\usepackage{amsmath}
\usepackage{graphicx}

\begin{document}

\title{Spiky density of states in large complex Al-Mn phases}

\author{G. Trambly de Laissardi\`ere \\
Laboratoire de Physique Th\'eorique et Mod\'elisation, \\ 
CNRS and Universit\'e de Cergy-Pontoise,
2 av. A. Chauvin, F-95302, Cergy-Pontoise, 
France.\\
Email: guy.trambly@u-cergy.fr}

\begin{abstract}
First-principle electronic structure calculations have been performed in crystalline complex phases $\mu$-Al$_{4.12}$Mn 
and $\lambda$-Al$_{4.6}$Mn using the TB-LMTO method. 
These atomic structures, related to quasicrystalline structures,
%of those phases are similar and both 
contain about 560 atoms in a large hexagonal unit cell. 
One of the main characteristic of their density of states is the presence 
of fine peaks the so-called ``spiky structure''.
From multiple-scattering calculations in real space, we show that these fine peaks are 
not artifacts in ab-initio calculations, since they result from a specific localization of electrons 
by atomic clusters of different length scales.
\end{abstract}

\maketitle

\section{Introduction}

Al based Quasicrystals, %\cite{shechtman84}, 
approximants and related complex
metallic phases reveal very unusual transport properties
(for review see Ref. \cite{Mayou07_revue} and 
Refs. therein).
Indeed, they exhibit a very high resistivity although the density of 
states (DOS) at the Fermi energy $E_{\rm F}$ is not very small.
This means that their transport properties are due mainly to a small diffusivity 
of electrons. 
Recently \cite{PRL06}, we proposed a new mechanism for the conductivity 
in those systems where the velocity of charge carriers is very low. 
This mechanism is related with a specific localization of electrons 
by complex atomic structures.
Several numerical studies argue that this localization leads to 
fine peaks in the DOS, called {\it ``spiky peaks''}.
Thus, spiky DOS 
were predicted in icosahedral small approximants
(for instance $\rm \alpha\,$Al-Mn-Si 1/1-approximant
\cite{Fujiwara89,Zijlstr03},
1/1\,Al-Cu-Fe \cite{GuyPRB94_AlCuFe}, 1/1\,Al-Pd-Mn
\cite{Krajvci95}).
In the case of crystalline approximants, it is obvious that
these peaks are associated to a
small electron velocity (flat dispersion
relations) 
\cite{Fujiwara89}.
Such properties are not specific 
of quasicrystals as they are also observed 
in many crystal related to quasicrystals,
therefore it does not come  from the long
range quasiperiodicity. 
%They are also associated
%with local and medium range atomic order
%that are related to quasiperiodicity.
Indeed, we have predicted \cite{GuyPRB97} 
that fine peaks in the DOS could come
from electron confinement by atomic
clusters characteristic %\cite{Gratias00}
of the quasiperiodicity.
This is not in contradiction with a Hume-Rothery mechanism 
for stabilizing these phases
because this
tendency to localization has a small effect on the
band energy \cite{GuyPRB97}.
The existence of spiky DOS in quasicrystals and approximants
with large unit cell is however still  much debated
experimentally (Refs. \cite{Dolinsek00,Stadnik01,Widmer06}  and references therein)
and theoretically
(Refs. \cite{GuyPRB97,Zijlstra00,Guy03} and references therein).
Very recently, R. Widmer et al. \cite{Widmer06}, found a clear signature of spiky local
DOS near Fermi level in scanning tunneling microscopy and spectroscopy 
at low temperature.

In this paper, 
we present ab-initio calculsations and multiple scattering calculations
that prove
the existence on fine peaks in the DOS
of complex Al-Mn crystals with large unit cell.

%\section{Atomic structures}
\section{First-principle study}

The $\mu$-$\rm Al_{4.12}Mn$ phase is crystallized in a hexagonal 
structure ($\rm P6_{3}/mmc$ space group) with large unit cell: $a=19.98(1)$\,$\rm \AA$
and $c=24.673(4)$\,$\rm \AA$ \cite{shoemaker89}. 
Each unit cell contains 110 Mn atoms and $\sim$453 Al atoms.
As explained in Ref. \cite{Duc03}, 
we have modified slightly  the  Wyckoff position  of one Al atom 
in order to avoid  too short interatomic distances.
The $\lambda$-$\rm Al_{4.6}Mn$ 
crystalizes also in $\rm P6_{3}/mmc$ 
with $a=23.382(9)$\,$\rm \AA$ and $c=12.389(2)$\,$\rm \AA$ \cite{Kreiner97}.
Its atomic arrangements
have many similarities with those of 
$\mu$-$\rm Al_{4.12}Mn$.

DOSs
are calculated 
by using the self-consistent tight-binding
linear muffin tin orbital (TB-LMTO)
method~\cite{andersen95}, 
in the atomic sphere approximation (ASA) without spin polarization.
The LMTO basis includes all angular momenta up to $l=2$.
%and the valence states are Al(3$s$,3$p$,3$d$)
%and Mn (4$s$,4$p$,3$d$).
The Brillouin zone (BZ) integration on the $k$ space was done by
the tetrahedron method with $N_k$ $k$-points in a BZ. 
LMTO DOS 
of $\mu$-$\rm Al_{4.12}Mn$ have been already calculated 
by D. Nguyen-Manh and G. Trambly de Laissardi\`ere \cite{Duc03}
for analyzing the origin of magnetism in complex aluminides related to
quasicrystals \cite{simonet98,Hippert99}.
Here, we focus on the shape of the DOS without spin-polarization. 
To increase the accuracy
of  calculation, we increase the number $N_k$.
In Ref.~\cite{Duc03} TB-LMTO calculation in $\mu$  phase
was performed with $N_k=9$, 
whereas $N_k=3375$ in actual calculation for $\mu$ phase,
and  $N_k=1458$, for  $\lambda$ phase.

\begin{figure}[t]
\begin{center}
\includegraphics[width=7cm]{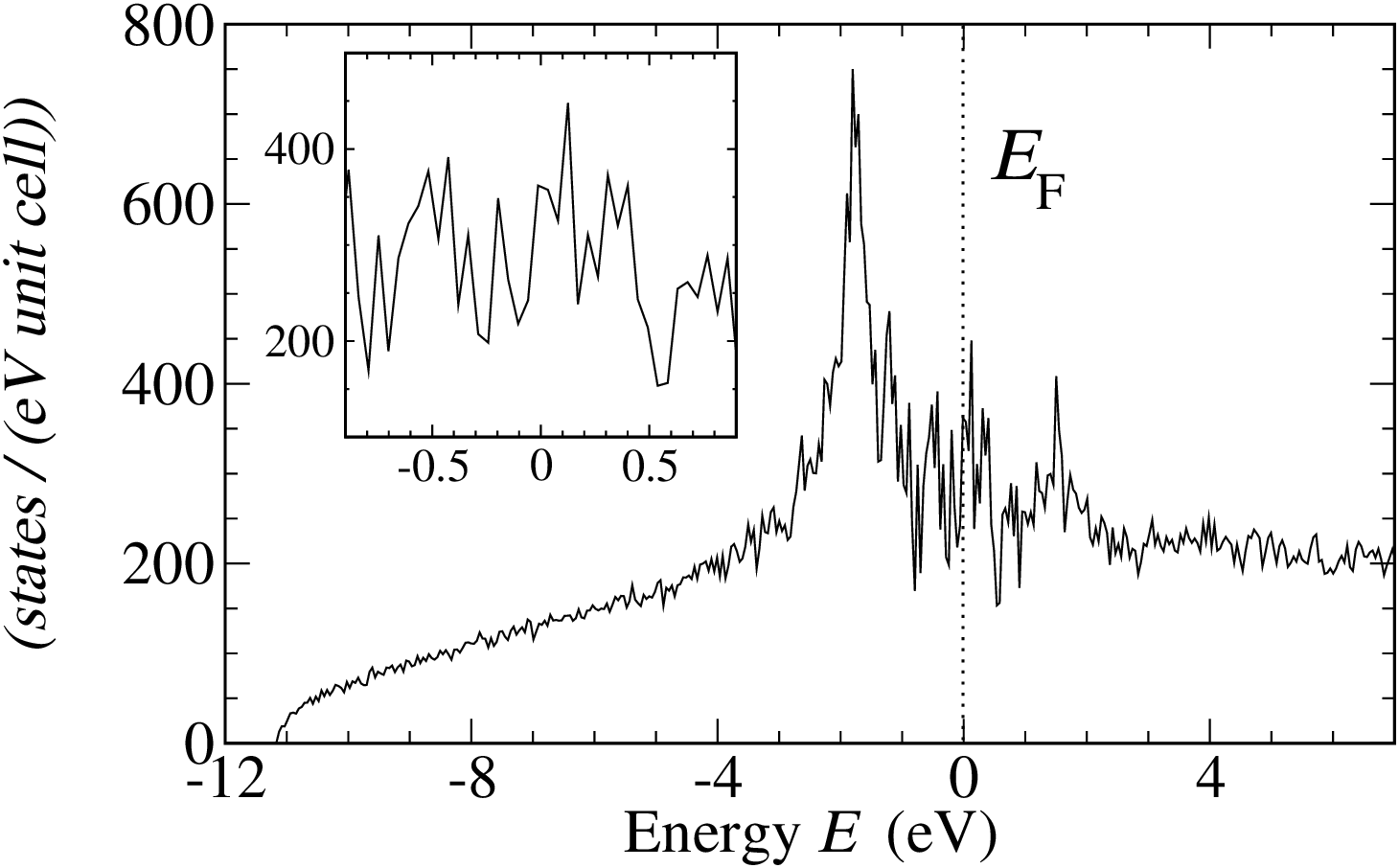}
\caption{Total LMTO DOS in $\mu$-$\rm Al_{4.12}Mn$.
Insert: Total DOS around $E=0$.
\label{Fig_mu_dos}
}
\end{center}
\end{figure}

%%%%%%%%%%%%%%%%%%%%%%%%%%%%%%%%%%%%%%%%%%%%%%%%%%%%%%%%%%%%%%%%%
\begin{figure}[t]
\begin{center}
\includegraphics[width=7cm]{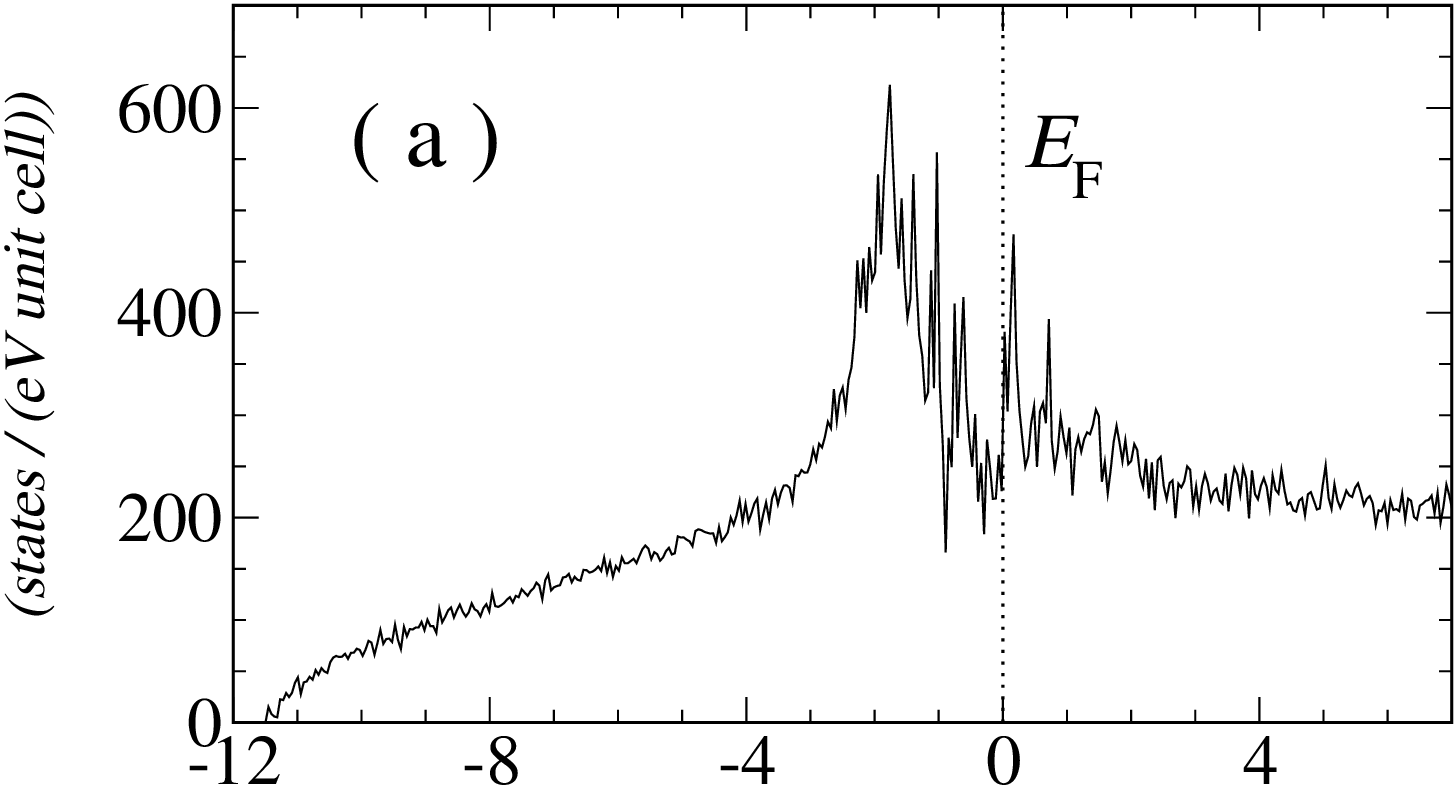}

\vskip .2cm
\includegraphics[width=7cm]{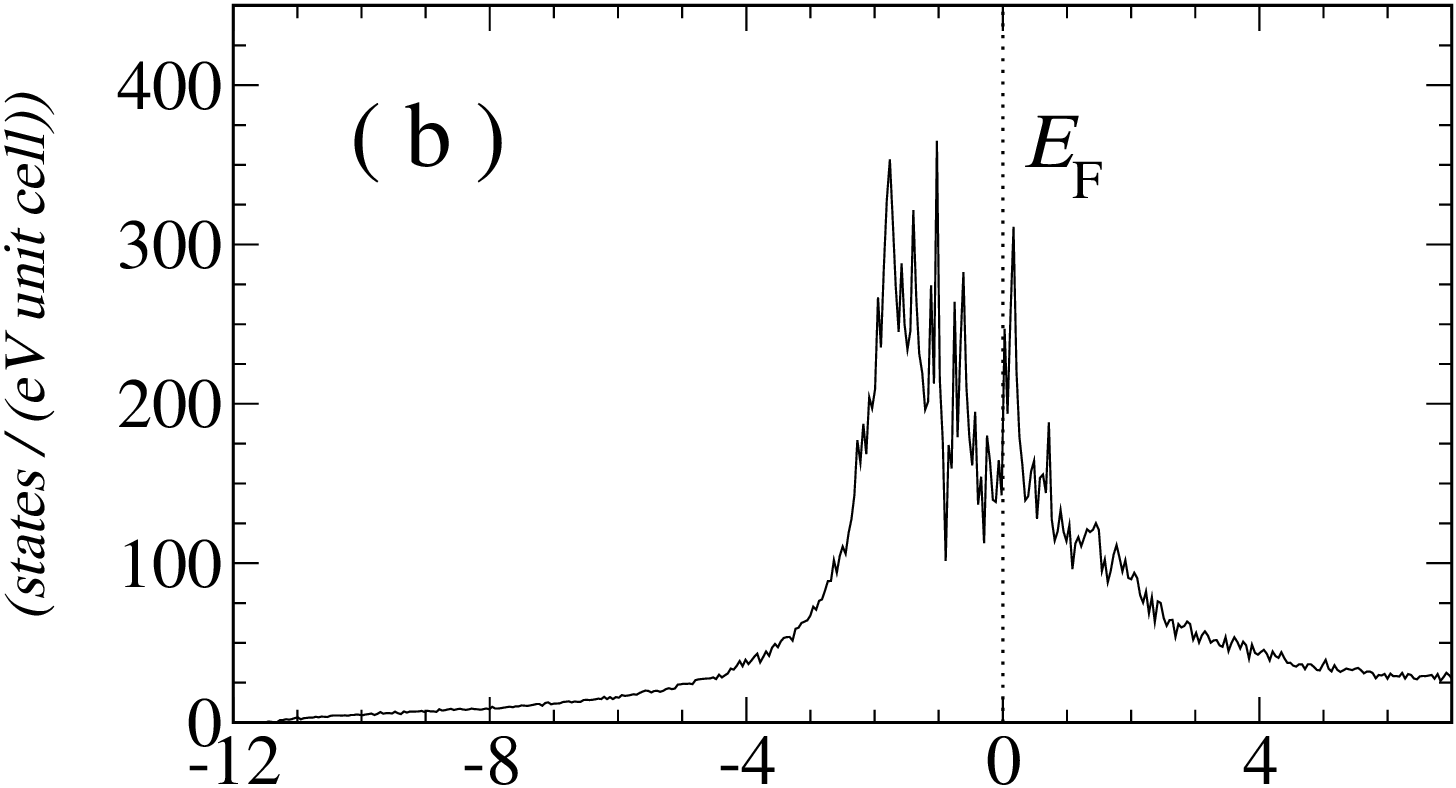}

\vskip .2cm
\includegraphics[width=7cm]{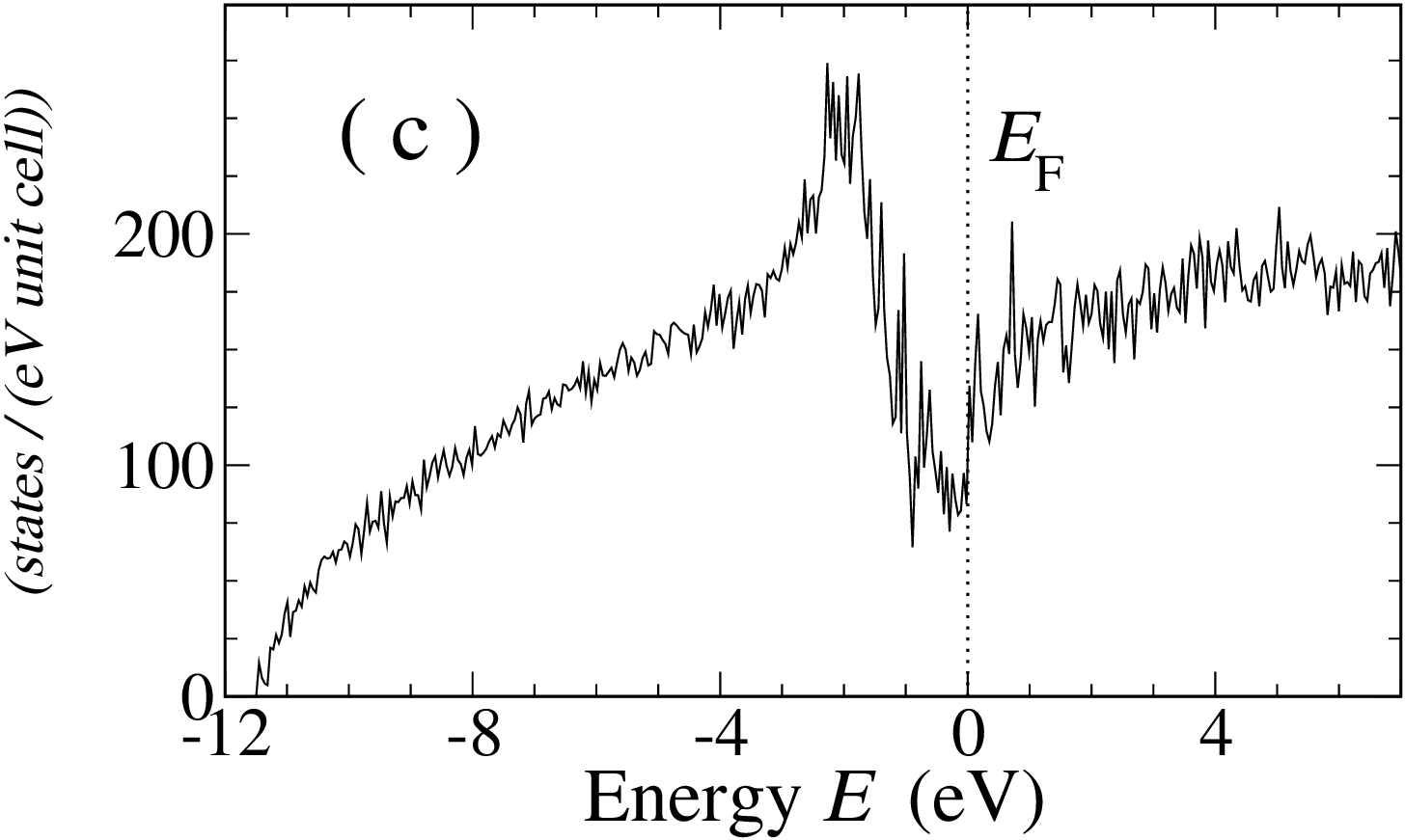}

\caption{LMTO DOS in $\lambda$-$\rm Al_{4.6}Mn$:
(a) Total DOS, (b) local Mn DOS
and (c) local Al DOS.
\label{Fig_lambda_dos}
}
\end{center}
\end{figure}

The total DOSs
in $\mu$ and $\lambda$ phases
are very similar
(figures~\ref{Fig_mu_dos} and \ref{Fig_lambda_dos}).
At low energies,
the parabola due to the
Al nearly-free states is clearly seen.
The large $d$ 
band (figure~\ref{Fig_lambda_dos}(b)) 
from $-2$ up to $1$\,eV
is due to a strong $sp$-$d$ hybridization 
\cite{Fujiwara89,Krajvci95,PMS05,dankhazi93}.
The sum of local DOSs on Al atoms, shown
in figure~\ref{Fig_lambda_dos}(c), is mainly $sp$ DOS.
As expected for a Hume-Rothery stabilization,
there is a wide pseudogap in $sp$ DOS
near $E_{\rm F}$ 
\cite{Fujiwara89,GuyPRB94_AlCuFe,Krajvci95,Hippert99,PMS05,dankhazi93}.
The pseudogap in the total DOS is narrower 
because the $d$ states of 
Mn atoms must fill up it partially \cite{PMS05}.
As shown on the insert figure~\ref{Fig_mu_dos},
DOS around $E_{\rm F}$ consists in a set of spiky peaks.
Nevertheless the resolution is not well
define in the TB-LMTO calculations \cite{Zijlstr03}, therefore these
calculations are not enough to prove the existence of spiky peaks.

\section{Spiky density of states ?}

%%%%%%%%%%%%%%%%%%%%%%%%%%%%%%%%%%%%%%%%%%%%%%%%%%%%%%%%%%%%%%%%%
\begin{figure}[t]
\begin{center}
\includegraphics[width=6cm]{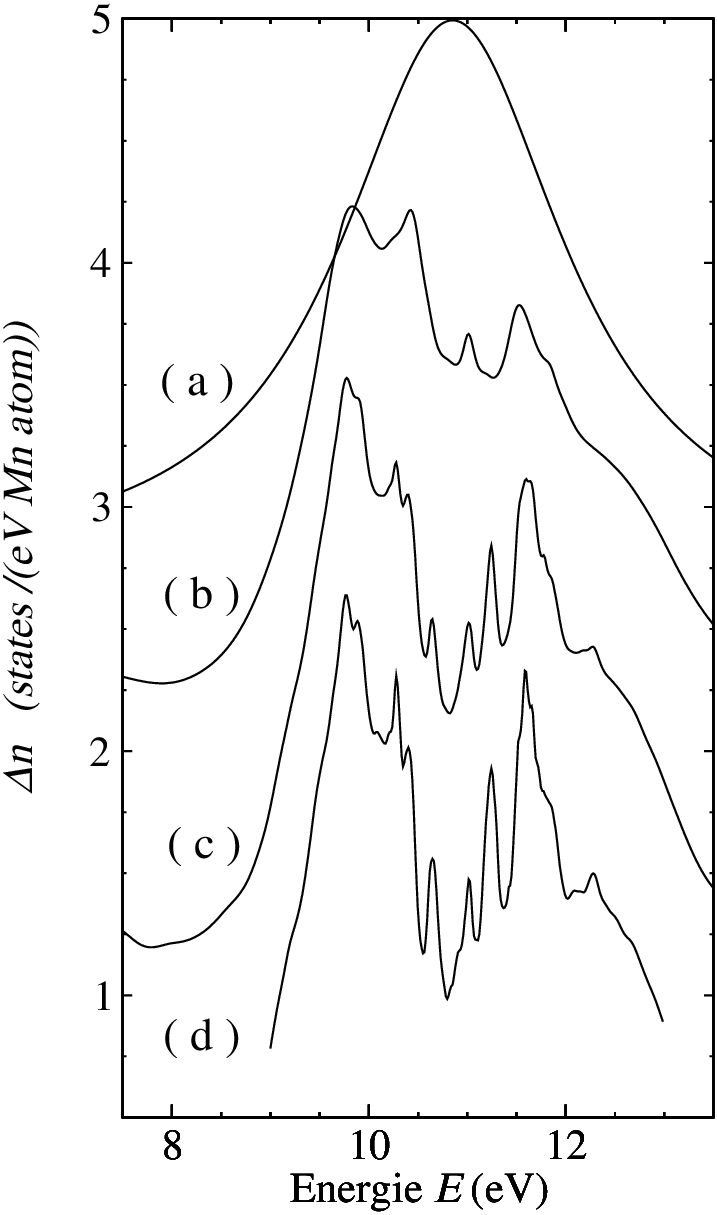}
%\vskip 2cm
\caption{Variation $\Delta n(E)$ of the total DOS due to $X$ Mn atoms in 
free electron matrix.
%simulating Al atoms. 
Mn atoms are located on the Mn sub-lattice of $\mu$-$\rm Al_{4.12}Mn$.
(a) $X=1$ is the Virtual Bound States.
(b) $X=110$ corresponds to 1 unit cell of $\mu$ phase.
(c) $X=880$, 8 unit cells of $\mu$ phase.
(d) $X=1980$, 18 unit cells of $\mu$ phase.
\label{Fig_cluster_mu}
}
\end{center}
\end{figure}

%%%%%%%%%%%%%%%%%%%%%%%%%%%%%%%%%%%%%%%%%%%%%%%%%%%%%%%%%%%%%%%%%
\begin{figure}[t]
\begin{center}
\includegraphics[width=6cm]{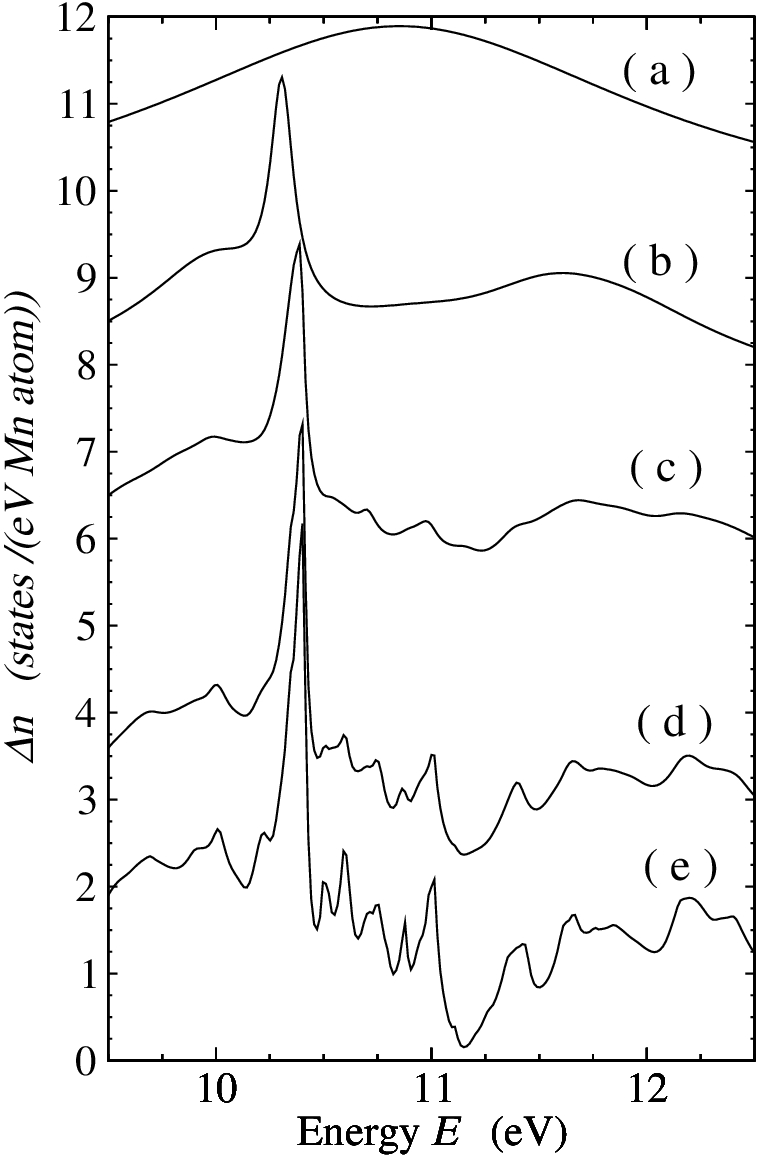}
%\vskip 2cm
\caption{Variation $\Delta n(E)$ of the total DOS due to $X$ Mn atoms in  
free electron matrix.
% simulating Al atoms.
Mn atoms are located on the Mn sub-lattice of $\alpha$-Al-Mn-Si cubic 1/1 approximant.
(a) $X=1$ is the Virtual Bound States.
(b) $X=12$, 1 Mn icosahedron.
(c) $X=192$, 16 Mn icosahedra on  bcc lattice.
%(like in $alpha$-AlMnSi approximant).
(d) $X=648$, 54 Mn icosahedra on  bcc lattice.
(e) $X=1536$, 128 Mn icosahedra on bcc lattice.
%The dashed line is $\Delta \rho$ value due to Mn-Mn pair interactions \cite{icq8}
%in the case (c).
\label{Fig_cluster_alpha}
}
\end{center}
\end{figure}

To prove the existence or not of fine peaks in DOS, 
we calculate the DOS for a large assembly of 
atoms by using a multiple scattering approach in real space.
Therefore there is no error due to Brillouin zone (BZ) integration as 
with TB-LMTO method.
Multiple scattering calculation
is not an ab-initio calculation, 
but it 
allows to discuss qualitatively the existence of spiky peaks.

In a standard description of intermetallic alloys,
one starts from a muffin-tin potential,
which varies in spheres centered on the atoms and
is constant in the space between the spheres.
As far as the band structure is concerned an atom is entirely
characterized by its scattering properties for incident plane waves.
%Mathematically, the central quantity is the {\bf T}-matrix.
%In the same way, 
The effect of a group of atoms on band structure
will be characterized by its {\bf T}-matrix,
that is by its scattering properties of plane
waves (Ref. \cite{GuyPRB97} and Refs. therein).
%Starting from the {\bf T}-matrix of various group of Mn atoms,
%We analyzed the scattering properties of 
%an assembly of  Mn atoms through
%a calculation of the variation $\Delta n(E)$
%of the total DOS due to the Mn atoms.
To simulate Al(Si)-Mn phases,
we consider only the potential
due to the Mn atoms, which are strong scatterers,
and we neglect the potential of all the Al (and Si) atoms
that are considered as weak scatterers.
%Furthermore, following a classical approximation
%we retain only the potential due to the $d$ orbital
%of the Mn atoms, which are strong scatterers.
%This means that in the framework of the scattering
%theory we consider only the phase shift for the 
%$l = 2$ components of the orbital moment.
%Thus in our model, 
Furthermore, %following a classical approximation,
we consider that
the Mn $d$ orbitals
are the only source of
scattering for the conduction electrons (i.e. the $sp$ electrons).
Within these approximations a  complex Al(Si)-Mn phase
becomes an assembly of $X$ atoms of Mn in a jellium 
simulating Al (Si) atoms (free $sp$ electron matrix).
The total DOS of this system is:
\begin{eqnarray}
n(E) = n^{free}_{sp}(E) + \Delta n(E),
\end{eqnarray}
where $n^{free}_{sp}$ is the DOS of free electrons
without $sp$-$d$ interaction,
and $\Delta n$, the variation of DOS due to the Mn atoms.
$\Delta n(E)$ is calculated by using the Lloyd formula
\cite{Gonis92,GuyPRB97}, which is
a generalization for several Mn atoms 
%in jellium
of the Friedel sum rule for
one  Mn atom in jellium:
\begin{eqnarray}
\Delta n(E) = - \frac{2}{\pi}{\rm Tr}
\left( {\bf M}(E)^{-1} \frac{{\rm d}}{{\rm d}E}{\bf M}(E)
\right).
\label{EquationLloyd}
\end{eqnarray}
The matrix elements of ${\bf M}(E)$
are calculated form the {\bf T}-matrix:
\begin{eqnarray}
{\bf M} = {\bf t}^{-1} - {\bf G}
\end{eqnarray}

where
${\bf G}$ is the propagator that coupled $d$ orbitals of different Mn scatters,
and
%%\begin{equation}
%%{\bf t} = 
%%\left( \begin{array}{cccc}
%%t_1 & & & \\
%% &  t_2 & & \\
%% &   & \ddots  & \\
%% &   & & t_X \end{array}
%%\right).
%%\end{equation}

\begin{equation}
{\bf t} = 
\left( \begin{array}{ccc}
t_1 &  & \\
 &    \ddots  & \\
 &    & t_X \end{array}
\right).
\end{equation}

$t_x$ is the element of the transfer matrix due to the Mn scatter indexed by $x$.
%and $X$ is the total number of Mn atoms.
$t_x$ depends %\cite{Gonis92} 
only on the phase shift $\delta_d$ in wavefunction of the $sp$ states
due to their scattering:

\begin{eqnarray}
\tan \delta_d(E)= \frac{- \Gamma}{E-E_d} 
\end{eqnarray}
with
\begin{eqnarray}
2\Gamma = \pi 
%|\langle d | {\bf k} \rangle |^2 
V^2
n^{free}_{sp}(E_{\rm F}).
\label{EqTanDelta}
\end{eqnarray}

where $2\Gamma$ is the  width of the $d$ resonance and
$E_d$ the on-site $d$ energy.
%\begin{eqnarray}
%2\Gamma = \pi 
%%|\langle d | {\bf k} \rangle |^2 
%V^2
%n^{free}_{sp}(E_{\rm F}).
%\end{eqnarray}
%$n^{free}_{sp}$ is the $sp$ DOS without $sp$-$d$ interaction
%and 
$V$ is the coupling between one $d$ orbital and $sp$ eigenstates \cite{PMS05}. 
When $X=1$ (impurity limit),
the
total $sp$ DOS, $n_{sp}$, is not modified by the $sp$-$d$ hybridization,
%($n_{sp} = n^{free}_{sp}$, compensation theorem), 
and the variation $\Delta n$ of the 
total DOS due to the Mn atom is a Lorentzian (Virtual Bound State)
centered on energy $E_d$ and with a  
width equals to $2\Gamma$.

To simulate complex Al(Si)-Mn,
% by and assembly of Mn atoms
%in a jellium (free electrons matrix) simulating Al (Si) atoms.
the only parameters are the values of $E_d$ and 
$\Gamma$, and positions of Mn atoms in the real space.
We use the following realistic values \cite{PMS05}: $E_d = 10.9$~eV and $2\Gamma = 3$~eV,
and atomic positions determined from X-ray refinements.
The DOS of several assemblies of Mn atoms are thus determined for Mn belonging to 
the Mn-sublattice of $\mu$-phase (figure \ref{Fig_cluster_mu})
and $\alpha$-AlMnSi 1/1-approximant (figure \ref{Fig_cluster_alpha}). 
%LMTO DOS of $\alpha$ phases
%is well known \cite{Fujiwara89,Zijlstr03}.
For  $\mu$ and $\alpha$ structures, fine peaks appear in the DOS
when a large number of Mn scatterers is 
taken into account (large value of $X$). 
Widths of these peaks are about $40-150$~meV which agrees
with ab-initio calculations 
(figures \ref{Fig_mu_dos} and \ref{Fig_lambda_dos} and Ref. \cite{PMS05}).
%As explain in Refs.~\cite{GuyPRB97}, 
Each fine peak is a signature
of states confined by atomic clusters,
so-called {\it ``Cluster Virtual Bound states''} \cite{GuyPRB97}.
When the number of Mn scatters increases, new peaks appear.
This means that atomic structure confines electrons on several 
length scales corresponding  to different sizes of clusters, respectively.

%\section{Conclusion}
Summarizing,
multiple-scattering calculations 
%by assemblies of atoms 
have been performed in real space
to simulate DOS of 
complex Al(rich)-Mn phases related to quasicrystals.
They show 
that a localization of electrons by large assemblies of atoms 
leading to fine peaks in the DOS.
This confirms the existence of the spiky DOS found by ab-initio calculations.

%\section*{Acknowledgments}
This work owes much to 
discusion with D. Mayou and D. Nguyen-Manh with whom ab-initio study 
of $\mu$-phase has been initiated.
The computations have been performed at the
Service Informatique Recherche (S.I.R.),
Universit\'e de Cergy-Pontoise.
%Part of the numerical results has been
%obtained by using  the Condor Project
%(http:$/\!/$www.condorproject.org$/$).
I thank also Y. Costes and B. Marir, S.I.R.,
for computing assistance.


\begin{thebibliography}{}

\bibitem{Mayou07_revue} Mayou, D.; Trambly de Laissardi\`ere, G.:
Quantum transport  in quasicrystals and complex metallic alloys.
In {\it Quasicrystals}, series ``Handbook of Metal Physics'',
(Eds T. Fujiwara, Y. Ishii), 
p. 209-265.
Elsevier Science, 2008.

\bibitem{PRL06}
Trambly de Laissardi\`ere, G.; Julien J. P.; Mayou, D.:
Quantum Transport of Slow Charge Carriers in Quasicrystals and Correlated Systems.
Phys. Rev. Lett.  {\bf 97}  (2006) 026601.
%\bibitem{ICQ9} G. Trambly de Laissardi\`ere, J. P. Julien and D. Mayou,
%Phil. Mag. {\bf 86} 663 (2006).

\bibitem{Fujiwara89} %DOS alpha AlMnSi\\
Fujiwara T.:
Electronic structure in the Al-Mn alloy crystalline analog of quasicrystals.
{Phys. Rev.}  B{\bf 40} (1989) 942-946.
%
%\bibitem{Fujiwara93} 
Fujiwara, T.; Yamamoto, S.; Trambly de
Laissardi\`ere, G.:
Band Structure Effects of Transport Properites in Icosahedral Quasicrystals.
{Phys. Rev. Lett.}  {\bf 71} (1993) 4166-4169.
%at. Sci. Forum  {\bf 150}-{\bf 151} 387 (1994).


\bibitem{Zijlstr03}
%\item[{\it S6.}] %alpha AlMnSi \\
Zijlstra, E. S.; Bose S. K.:
Detailed ab initio electronic structure study of two approximants
to Al-Mn based icosahedral quasicrystals.
{Phys. Rev.} B{\bf 67}, (2003) 224204.


\bibitem{GuyPRB94_AlCuFe}
%DOS et Conductivite i- approximant AlCuFe\\
Trambly de Laissardi\`ere, G.;  Fujiwara, T.:
Electronic structure and conductivity in a model approximant
of the icosaheral quasicrystals Al-Cu-Fe.
{Phys. Rev.}   B{\bf 50} (1994) 5999-6005 .

\bibitem{Krajvci95} 
Kraj\v{c}\'{\i}, M., 
{\it et al.}:
%Windisch M, Hafner J, Kresse G,
%Mihalkovic M 1995
%ATOMIC AND ELECTRONIC-STRUCTURE OF ICOSAHEDRAL AL-PD-MN ALLOYS AND APPROXIMANT PHASES
Atomic and electronic structure of icosahedral Al-Pd-Mn alloys and approximant phases.
{ Phys. Rev.} B {\bf 51} (1995) 17355-17378.

\bibitem{GuyPRB97} 
Trambly de Laissardi\`ere, G.; Mayou, D.:
Clusters and localization of electrons in quasicrystals 
{Phys. Rev. B}  {\bf 55} (1997) 2890-2893.
%
%\bibitem{GuyICQ6} G. 
Trambly de Laissardi\`ere, G.; Roche, S.; Mayou, D.:
%Electronic confinement by clusters in quasicrystals and approximants 
{Mat. Sci. Eng. A} {\bf 226}-{\bf 228} 986-989 (1997).

%\bibitem{Gratias00} %Atomic cluster in i- F type QC\\
%Gratias, D.; Puyraimond, F.; Quiquandon, M.; Katz, A.:
%Atomic clusters in icosahedral F-type quasicrystals
%{Phys. Rev.}  B{\bf 63 } (2000) 24202.

\bibitem{Dolinsek00} 
Dolin$\rm \check{s}$ek J, 
{\it et al.}:
%Klanj$\rm \check{s}$ek M,
%Apih T, Smontara A, Lasjaunias J C, Dubois J M,
%Poon S J 2000
Searching for sharp features in the pseudogap of icosahedral quasicrystals by NMR.
{\it Phys. Rev.} B{\bf 62} (2000) 8862-8870.

\bibitem{Stadnik01} %photoemission i Al-Pd-Mn
Stadnik, Z. M.,
{\it et al.}:
%Purdie D, Baer Y, Lograsso T A 2001
Absence of fine structure in the photoemission spectrum of the icosahedral Al-Pd-Mn quasicrystal.
{\it Phys. Rev.} B{\bf 64} (2001) 214202.

\bibitem{Widmer06} 
Widmer, R., 
{\it et al.}:
Low-temperature scanning tunneling spectroscopy on the 5-fold surface of the icosahedral AlPdMn quasicrystal.
Phil.Mag. {\bf 86} (2006) 781-787.
Maeder, R.,  {\it et al.}: {\it private communication}.

\bibitem{Zijlstra00}
%Non-spiky density of states of an icosahedral QC\\
Zijlstra, E. S.; Janssen, T.: 
Non-spiky density of states of an icosahedral quasicrystal
{Europhys. Lett.} {\bf 52} (2000) 578-583.

\bibitem{Guy03} Trambly de Laissardi\`ere, G.:
Interplay between electronic structure
and medium range atomic order in hexagonal
$\rm \beta\,Al_9Mn_3Si$ and $\rm \varphi\,Al_{10}Mn_3$ crystals.
Phys. Rev. B  {\bf 68} (2003) 045117.

\bibitem{shoemaker89}
Shoemaker, C. B.; Keszler, D.; Shoemaker, D.:
Structure of $\mu$-MnAl$_4$ with Composition Close to that of Quasicrystal Phases.
Acta Cryst. B{\bf 45}, (1989) 13-20.

\bibitem{Duc03} 
Nguyen-Manh D.; Trambly de Laissardi\`ere, G.:
First-principles predictions of magnetic properties for
a complex and strongly related to quasicrystalline
phase: $\mu$-Al$_4$Mn.
\textit{J. Mag. Mag. Mater.}, {\bf 262}  (2003) 496-501.

\bibitem{Kreiner97}
%Structure de la phase lambda-Al4Mn \\
Kreiner, G; Franzen, H.F.: 
The crystal structure of $\lambda$-Al$_{4.6}$Mn 
{\it J. Alloys Comp.} {\bf 261} (1997) 83-104.

\bibitem{andersen95}
Krier, G.; Jepsen, O.; Burkhart A.; Andersen, O. K.:
The TB-LMTO program. Stuttgart, (1995).

\bibitem{simonet98}
Simonet, V.,
{\it et al.}:
Origin of magnetism in crystalline and quasicrystalline AlMn and AlPdMn phases.
%F. Hippert, M. Audier and G. Trambly de Laissardi\`ere, 
Phys. Rev. B{\bf 58} (1998) R8865-R8868.

\bibitem{Hippert99} 
Hippert, F., 
{\it et al.}:
Magnetic properties of AlPdMn approximant phases 
%V. Simonet,
%G. Trambly de Laissardi\`ere, M. Audier and Y. Calvayrac,
J. Phys: Condens. Matter  {\bf 11} (1999) 10419-10435.
Trambly de Laissardi\`ere, G.;  Mayou, D.:
Magnetism in Al(Si)-Mn quasicrystals and related phases 
Phys. Rev. Lett. {\bf 85} (2000) 3273-3276.
Pr\'ejean, J. J.; Hippert, F.: 10th International Conference on Quasicrystals,
Zurick, 2008.

\bibitem{PMS05}
%\item[11.]
Trambly de Laissardi\`ere, G.; Nguyen-Manh D.; Mayou, D.:
Electronic structure of complex Hume-Rothery phases and quasicrystals
in transition metal aluminides.
{Prog.  Mater. Sci.}  {\bf 50} (2005) 679-788.

\bibitem{dankhazi93}
Z. Dankhazi, Z.,
{\it et al.}:
%G. Trambly de Laissardi\`ere, D. Nguyen-Manh, 
%E. Berlin and D. Mayou, 
%THEORETICAL AND EXPERIMENTAL ELECTRONIC DISTRIBUTIONS IN AL6MN 
Theoretical and experimental electronic distribution in Al$_6$Mn.
J. Phys. Condens. Matter, 
{\bf 5} (1993) 3339-3350. 
Trambly de Laissardi\`ere, G., 
{\it et al.}:
% EXPERIMENTAL AND THEORETICAL ELECTRONIC DISTRIBUTIONS IN AL-CU-BASED ALLOYS
Experimental and theoretical electronic distributions in Al-Cu-based alloys.
Phys. Rev. B{\bf 51} (1995) 14035-14047.
Belin, E; Mayou, D.:
% ELECTRONIC-PROPERTIES OF QUASI-CRYSTALS 
Electronic properties of quasi-crystals.
Phys. Scr. T {\bf 49} (1993) 356-359.
%G. Trambly de Laissardi\`ere,
%Z. Dankh\'azi, E. Belin, A.  Sadoc, D. Nguyen--Manh, D. Mayou, M. A. Keegan and D. Papaconstantopoulos,
%{Phys. Rev B} {\bf 51} 14035 (1995).

\bibitem{Gonis92} Gonis A.:
Green Functions for Ordered and
Disordered Systems.
(Ets: E. van Groesen and E.M. de Jager)
Studies in Mathematical Physics,
vol. 4. North Holland, Amsterdam 1992.

\end{thebibliography}
\end{document}